\documentclass[journal]{IEEEtran}

\usepackage{cite}
\ifCLASSINFOpdf
\else
\fi
\usepackage{multirow}

\usepackage[cmex10]{amsmath}

\usepackage{algorithm}
\usepackage{algorithmic}
\usepackage{setspace}

\usepackage{array}
\usepackage{graphicx}
\usepackage{caption}
\usepackage{subcaption}
\usepackage{color}

\usepackage{stfloats}
\usepackage{amsmath,amsfonts,amssymb,amscd}
\usepackage{mathrsfs}
\usepackage{graphicx,epsfig}

\usepackage{cases}

\usepackage[font=singlespacing]{caption}
\usepackage{lscape}
\usepackage{verbatim}
\usepackage{epstopdf}
\usepackage{balance}

\usepackage{graphicx}

\usepackage{bbm}

\newcommand{\bR}{\mathbb{R}}

\newcommand{\T}{\!^\top\!}

\newcommand{\xh}{\hat{x}}

\newcommand{\xhi}{\xh_i}
\newcommand{\li}{\kappa_i}
\newcommand{\lj}{\kappa_j}
\newcommand{\ai}{\ell_i}

\newcommand{\xhik}{\xhi(k)}

\newcommand{\ljk}{\lj(k)}
\newcommand{\aik}{\ai(k)}

\newcommand{\xikp}{x_i(k+1)}

\newcommand{\xhikp}{\xhi(k+1)}

\newcommand{\likp}{\li(k+1)}

\DeclareSymbolFont{matha}{OML}{txmi}{m}{it}% txfonts
\DeclareMathSymbol{\varv}{\mathord}{matha}{118}
\begin{document}
\title{Distributed Vehicle Grid Integration Over Communication and Physical Networks}


\author{\IEEEauthorblockA{\large{Dimitra~Apostolopoulou, Rahmat~Poudineh, and Anupama~Sen}}

\thanks{D. Apostolopoulou is with the Department of Electrical and Electronic Engineering at City, University of London, London, UK EC1V 0HB. E-mail: \texttt{Dimitra.Apostolopoulou@city.ac.uk}

R. Poudineh and A. Sen are with the Oxford Institute for Energy Studies, Oxford, UK OX2 6FA. E-mail: \texttt{\{Rahmat.Poudineh,Anupama.Sen\}@oxfordenergy.org}}}


\maketitle

\begin{abstract}

This paper proposes a distributed framework for vehicle grid integration (VGI) taking into account the communication and physical networks. To this end, we model the electric vehicle (EV) behaviour that includes time of departure, time of arrival, state of charge, required energy, and its objectives, e.g., avoid battery degradation. Next, we formulate the centralised day ahead distribution market (DADM) which explicitly represents the physical system, supports unbalanced three phase networks with delta and wye connections, and incorporates the charging needs of EVs. The solution of the centralised market requires knowledge of EV information in terms of desired energy, departure and arrival times that EV owners are reluctant in providing. Moreover, the computational effort required to solve the DADM in cases of numerous EVs is very intensive. As such, we propose a distributed solution of the DADM clearing mechanism over a time-varying communication network. We illustrate the proposed VGI framework through the 13-bus, 33-bus, and 141-bus distribution feeders.

\end{abstract}



% \fi
\IEEEpeerreviewmaketitle

\section{Introduction}

A rapid increase in the adoption of electric vehicles (EVs) has been recorded worldwide in the last years. This is partly due to emissions reduction goals, e.g., the UK has committed to reduce its emissions by 80\% by 2050; and the decreasing prices of lithium ion batteries. However, there are several obstacles that need to be surpassed in order to promote a vast adaptation of EVs. In particular, the power consumption of typical household appliances, e.g., washing machines, refrigerators, is very different compared to that of an EV charger. It has been shown in~\cite{CROZIER2018474} that an uncontrolled EV charging scheme would increase the peak national demand in the UK by 20 GW assuming all vehicles were electrified. Moreover, the after diversity maximum demand in distribution systems at a house level increases by 1 kW with the integration of an EV~\cite{electric_avenue}. These effects and the associated costs with upgrading the electric power network (e.g.,~\cite{rep1}) may be mitigated if instead of uncontrolled charging, active charging techniques are used. In unidirectional active charging, EVs can modulate the charging power; and in the bidirectional case, EVs can also inject power back to the grid. We refer to the first one as Vehicle Grid Integration (VGI) and the latter is known as Vehicle to Grid (V2G)~\cite{SHI2018}. VGI may be seen as an intermediate solution between uncontrolled charging and V2G that requires less communication infrastructure. An additional benefit of VGI is that it is more considerate towards the EV battery compared to V2G. Under a VGI framework EVs may offer services to the transmission system operator, distribution system operator and facilitate the integration of renewable resources. For instance, EVs can offer peak shaving, power losses reduction, voltage regulation and frequency control in distribution systems (see, e.g.,~\cite{6663724, 7452714, 7286133}). The services that may be provided by an individual EV owner are small when compared to the large scale complex power system. However, an EV aggregator, who is in charge of operating the charging schedule of numerous EVs, may offer services to the grid ranging from kWs to MWs. 

In this paper we present a framework for a network-aware distributed VGI. More specifically, we propose a methodology to coordinate the services and operational constraints of three entities: the EV owner, the EV aggregator and the Distribution system operator (DSO). Each of these players has a different objective and privacy concerns. To this end, we formulate an optimisation problem that represents the day ahead distribution market (DADM) where the EVs and the aggregator participate. We explicitly model the EV behavior in terms of time of arrival and departure, required charging energy, and its objectives as well as the network constraints of an unbalanced three-phase distribution system. A centralised structure of the DADM by the DSO requires knowledge of EV information that EV owners are reluctant in providing. In this regard, we propose a distributed solution of the DADM clearing mechanism where the entities exchange limited insensitive information over a time-varying communication network, as is in reality, until they reach consensus. One challenge in this setting is that the aggregator would not be certain of the number of EVs available at a certain time instant, their state of charge and available energy, their arrival and departure times. As such we propose a methodology to provide the aggregators with a given confidence on the amount of capacity that they have available to participate in the market. In the numerical results section, we use the proposed framework in large-scale distribution feeders using realistic data for the EV behaviour and demonstrate its applicability and how it may be used by aggregators as well as DSOs for the system benefit.

Next, we discuss some relevant works in the literature which have also looked at the smooth integration of EVs in power systems. A thorough review of EV management schemes is given in~\cite{HU20161207}. Some studies have focused on modelling and control problems of EVs for valley-filling (e.g.,~\cite{6313962}), frequency regulation (e.g.,~\cite{7268776}), and facilitating the integration of renewable resources (e.g.,~\cite{RICHARDSON2013247}). Charging algorithms are categorised into two broad classes: centralised and decentralised approaches. In~\cite{7098444} a framework for centralised real-time EV charging management from an EV aggregator that participates in the energy and regulation markets is developed. A decentralised algorithm to optimally schedule EV charging that fills the valleys in electric load profiles is given in~\cite{6313962}. The authors design an algorithm that only requires each EV solving its local problem, hence its implementation requires low computation capability. In~\cite{7268776} the authors propose a distributed EV charging coordination mechanism to meet the daily mobility energy requirement of an EV fleet with respect to the day-ahead schedule of the EV aggregator and to meet the regulation dispatch signals sent to the EV aggregator by the system. However in the formulations above the network effects are neglected. In~\cite{8118112} a decentralised EV charging control scheme to achieve ``valley-filling'' while meeting heterogeneous individual charging requirements and satisfying distribution network constraints is proposed. However, the network formulation is based on a simplified representation that is not sufficiently accurate as shown in~\cite{8260205}. In~\cite{6663724} the authors propose a coordination methodology for the operation of EV owners, EV fleet operators and the DSO by only considering the cost minimisation subject to aggregated capacity to approximate the network effects.

 The remainder of the paper is organised as follows. In Section~\ref{sec:EVmod}, we describe the EV modelling, the EV objectives and constraints associated with charging decisions. In Section~\ref{sec:dadm}, we introduce the network modelling and define the DADM clearing problem taking into account the EV charging needs and objectives. In Section~\ref{sec:dadm_dis}, we formulate a distributed solution over a time-varying communication network to the DADM that addresses privacy and computational issues of the market participants. In Section~\ref{sec:unc_mod}, we propose a methodology to provide the EV aggregators with a given confidence on the amount of capacity that they have available to participate in the DADM. In Section~\ref{sec:num}, we illustrate the proposed VGI framework through the 13-bus, 33-bus, and 141-bus distribution feeders. In Section~\ref{sec:conc}, we summarize the results and make some concluding remarks.

\section{EV Modelling}

\label{sec:EVmod}

In this section, we describe the EV modelling, as well as the objectives and constraints associated with EV charing decision making.  The principal sources of uncertainty for an EV are (i) the time intervals that an EV is connected to the grid; (ii) the distances traveled by an EV, i.e., the amount of energy consumed from the battery due to driving; and (iii) the state of charge (SOC) of an EV at any point in time~\cite{GUILLE20094379}. We consider a collection of $E$ EVs denoted by the set $\mathscr{E} = \{ 1,2,\dots,E\}$ and a study period of $\mathscr{T} = \{1,\dots,T\}$ with $T$ intervals of size $\Delta t$. We assume that the EVs are connected at a three-phase network with $N_\text{bus}$ nodes denoted by the sets $\mathscr{N}_\text{bus} = \{1,\dots, N_\text{bus}\}$ and phases $\Phi = \{a,b,c\}$. In order to determine the location of the EVs, we need to determine the node and the phase that they are connected to. To this end, we define for each EV $j \in \mathscr{E}$ the triplet $\mathscr{H}_j = \{n_j,\phi_j,\xi_j\}$, where $n_j \in \mathscr{N}_\text{bus}$ is the node that the EV is connected to, $\phi_j \in \Phi$ the phase, and $\xi_j$ the type of connection which takes values delta or wye. For the entire set of vehicles $\mathscr{E}$ we define the collection of triplets $\mathscr{H} = \{\mathscr{H}_1,\dots, \mathscr{H}_E\}$. We introduce the energy consumed by EV $j$ for commuting at period $\mathscr{T}$ by $e_{j}$, $j \in \mathscr{E}$. We denote by $\mathscr{T}^{\text{dep}}_{j} = \{t^{\text{dep}}_{j,1}, \dots ,t^{\text{dep}}_{j,D_j}\}$ the set that indices the times EV $j$ departed within period $\mathscr{T}$ from home, where $D_j$ is the number of times that car $j$ departs from the house in $\mathscr{T}$.  $\mathscr{T}^{\text{arr}}_{j}= \{t^{\text{arr}}_{j,1}, \dots ,t^{\text{arr}}_{j,D_j}\}$ is the set that defines the times EV $j$ arrived at period $\mathscr{T}$ at home. We denote by $\pi_{j}(t)$ the availability of EV $i$ at time $t$ by:
\begin{equation}
\pi_{j}(t) = \left \{ \begin{matrix} 1, &0\leq t<t^{\text{dep}}_{j,1}, \\
0, & t^{\text{dep}}_{j,1} \leq t < t^{\text{arr}}_{j,1}, \\
1 , & t^{\text{arr}}_{j,1} \leq t < t^{\text{dep}}_{j,2}, \\
\vdots & \\
0 , & t^{\text{dep}}_{j,D_j} \leq t < t^{\text{arr}}_{j,D_j}, \\
1 , & t^{\text{arr}}_{j,D_j} \leq t < T .
 \end{matrix} \right.
\end{equation}
In this work, we only consider home charging however, the proposed framework can easily be expanded to include work and other public space charging. We denote by $y_{j}(t)$ the charging power of vehicle $j$ at time interval $t$. 

The charging constraints associated with the charging variables are the following:
\begin{equation}
\sum_{t \in \mathscr{T}} \pi_{j}(t) y_j(t) \Delta t = e_j, \forall j \in \mathscr{E},
\label{eq2}
\end{equation}
\noindent which ensures that each vehicle has received the right amount of energy at the end of the time horizon. The initial and final SOC are implicitly represented in \eqref{eq2} by appropriately defining $e_j$, for $j \in \mathscr{E}$. There are limits associated with each charging power which can be expressed as follows:
\begin{equation}
0 \leq y_j(t) \leq \pi_{j}(t) y_j^\text{max}(t) ,
\label{eq1}
\end{equation}
\noindent where $y_j^\text{max} (t)$ is the maximum value, e.g., $3.7$ kW for slow charging. Equation \eqref{eq1} ensures that at times when the EV $j$ is not available for charging $y_j(t)$ will be zero. 

The degradation cost of the EV battery is taken into account by minimising the 
second order polynomial of the charging rates~\cite{7039255}:
\begin{equation}
\sum_{t \in \mathscr{T}} \sum_{j \in \mathscr{E}} y_j^2(t).
\label{eq5}
\end{equation}

In this paper, we do not consider EVs as curtailable price responsive loads; thus we do not include a cost component for the charging power $y_j(t)$. As a result, the amount of energy necessary $e_i$ is pre-defined and stays constant. However, the EVs charge at the minimum possible cost of $e_j$ total energy due to the formulation of the day-ahead distribution market, which is formulated in Section~\ref{sec:DADM}. In future work, we will address the willingness of EV owners to participate in the market and modifying their desired energy $e_j$ based on price signals; thus making the energy $e_j$ a decision variable of the framework.

\section{Centralised Day-Ahead Distribution Market (DADM) Clearing Problem}

\label{sec:dadm}

We follow the DADMs model as described in~\cite{7862921}. In this section, we first introduce the network modelling and define the DADM clearing problem. We solve the market clearing for the period $\mathscr{T}$ and find the distribution location marginal prices (DLMPs); the real and reactive power quantities consumed or produced at each point in the network, so as to minimise the distribution network operator's cost minus the distributed participant benefits subject to linearised power flow relations and voltage magnitude constraints. We also model distributed generation (DG), such as photovoltaic (PV) resources whose capacity can be used for reactive power compensation and voltage control. 

\subsection{Network Modelling}

To reduce the computational complexity, a linear model is used for the modelling of three phase unbalanced networks, as described in~\cite{8260205}. The authors have validated its accuracy compared to a full AC power flow. Let us assume that the system has $N_\text{bus}$ three-phase buses denoted by the sets $\mathscr{N}_\text{bus} = \{1,\dots, N_\text{bus}\}$ and the phases $\Phi = \{a,b,c\}$; and $\ell$ lines denoted by the set $\mathscr{L} = \{1, \dots, \ell\}$. We denote by $Y \in \mathbb{C}^{3 N_\text{bus} \times 3 N_\text{bus}}$ the admittance matrix; by $s^Y  \in \mathbb{C}^{3N_\text{bus}}$ ($s^\Delta  \in \mathbb{C}^{3N_\text{bus}}$) the phase to line (phase to phase) complex power injections at each bus and $\varv \in  \mathbb{R}^{3N_\text{bus}}$ the magnitude of the bus complex voltages. We assume node 0 is the slack bus and partition the admittance matrix and the voltage magnitude vector as following $Y = \begin{bmatrix}
Y_{00} & Y_{0L} \\
Y_{L0} & Y_{LL}
\end{bmatrix}$, where $Y_{00} \in \mathbb{C}^{3 \times 3}$, $Y_{L0} \in \mathbb{C}^{3(N_\text{bus}-1)   \times 3}$, $Y_{0L} \in \mathbb{C}^{ 3 \times 3(N_\text{bus}-1) }$, and $Y_{LL} \in \mathbb{C}^{ 3(N_\text{bus}-1) \times 3(N_\text{bus}-1) }$; and  $\varv = [\varv_{0} ,\varv_{L}]^\top$ where $\varv_0 \in \mathbb{R}^3$ is the slack bus voltage magnitude and  $\varv_L \in \mathbb{R}^{3(N_\text{bus}-1)}$ the voltage magnitudes at remaining buses. Let us assume that the real (reactive) power phase to line injections are denoted by $p^Y \in \mathbb{C}^{3(N_\text{bus}-1)}$ ($q^Y \in \mathbb{C}^{3(N_\text{bus}-1)}$) and the real (reactive) power phase to line load is denoted by $p^Y_d \in \mathbb{C}^{3(N_\text{bus}-1)}$ ($q^Y_d \in \mathbb{C}^{3(N_\text{bus}-1)}$) for all buses than the slack bus, i.e., $\forall$ $n \in \mathscr{N}_\text{bus}/\{0\}$. The real (reactive) power phase to phase injections are denoted by $p^\Delta \in \mathbb{C}^{3(N_\text{bus}-1)}$ ($q^\Delta \in \mathbb{C}^{3(N_\text{bus}-1)}$) and the real (reactive) power phase to phase load is denoted by $p^\Delta_d \in \mathbb{C}^{3(N_\text{bus}-1)}$ ($q^\Delta_d \in \mathbb{C}^{3(N_\text{bus}-1)}$) for all buses than the slack bus, i.e., $\forall$ $n \in \mathscr{N}_\text{bus}/\{0\}$. 

The fixed-point linearisation around a nominal point $(\hat{s}^Y, \hat{s}^\Delta, \hat{\varv})$ renders the following relationships for the network representation:
\begin{equation}
\varv = K^Y \begin{bmatrix} p^Y-p^Y_{d} \\ q^Y-q^Y_d \end{bmatrix}+K^\Delta \begin{bmatrix} p^\Delta-p^\Delta_{d} \\ q^\Delta-q^\Delta_d \end{bmatrix} + b,
\label{eq3}
\end{equation}
where  $K^Y =  \text{diag}(h)\text{Re}(\text{diag}(h)^{-1} M^Y)$, $K^\Delta =  \text{diag}(h)\text{Re}(\text{diag}(h)^{-1} M^\Delta)$, $b = |h|$, with 
\begin{equation*}
M^Y = \begin{bmatrix} 0_{3 \times 3(N_\text{bus}-1)} &0_{3 \times 3(N_\text{bus}-1)} \\ Y_{LL}^{-1} \text{diag}(\overline{\hat{\varv}}_L)^{-1} & -jY_{LL}^{-1} \text{diag}(\overline{\hat{\varv}}_L)^{-1} \end{bmatrix},
\end{equation*}
\begin{equation*}
M^\Delta= \begin{bmatrix} 0_{3 \times 3(N_\text{bus}-1)} &0_{3 \times 3(N_\text{bus}-1)} \\ Y_{LL}^{-1} H^T \text{diag}(H\overline{\hat{\varv}}_L)^{-1} & -jY_{LL}^{-1} H^T \text{diag}(H\overline{\hat{\varv}}_L)^{-1} \end{bmatrix},
\end{equation*}
and $h = \begin{bmatrix} \hat{\varv}_0 \\ -Y_{LL}^{-1} Y_{L0} \hat{\varv}_0 \end{bmatrix}$, where $\text{Re}(\cdot)$ denotes the real part of a complex number and $\overline{(\cdot)}$ its conjugate. The complex power at the substation denoted by $s_0 = p_0+jq_0 \in \mathbb{C}^{3}$ is given by:
\begin{equation}
s_0 = G^Y  \begin{bmatrix} p^Y-p^Y_{d} \\ q^Y-q^Y_d \end{bmatrix}+G^\Delta  \begin{bmatrix} p^\Delta-p^\Delta_{d} \\ q^\Delta-q^\Delta_d \end{bmatrix}  +c,
\label{eq4}
\end{equation}
\noindent where $G^Y = \text{diag}(\hat{\varv}_0) \overline{Y}_{0L} \overline{M}^Y$, $G^\Delta = \text{diag}(\hat{\varv}_0) \overline{Y}_{0L}\overline{M}^\Delta$ and $c = \text{diag}(\hat{\varv}_0) \left(\overline{Y}_{00} \overline{\hat{\varv}}_0-\overline{Y}_{0L} \overline{Y}_{LL}^{-1} \overline{Y}_{L0} \overline{\hat{\varv}}_0\right)$.

\subsection{Day-Ahead Distribution Market Clearing Formulation}

\label{sec:DADM}

The constraints associated with the DADM are the network constraints given by the  linearised load flow relationships in \eqref{eq3}-\eqref{eq4}, modified to include the charging variables $y_j(t)$, $\forall j \in \mathscr{E}, t \in \mathscr{T}$ as loads. Thus we have:
\begin{eqnarray}
&\varv(t) =& K^Y \begin{bmatrix} p^Y(t)-p^Y_{d}(t)-\tilde{y}^Y(t) \\ q^Y(t)-q^Y_d(t) \end{bmatrix} +\nonumber \\
&&K^\Delta \begin{bmatrix} p^\Delta(t)-p^\Delta_{d}(t)-\tilde{y}^\Delta(t) \\ q^\Delta(t)-q^\Delta_d(t) \end{bmatrix} + b, \forall t \in \mathscr{T},\qquad
\label{eq10}
\end{eqnarray}
and
\begin{eqnarray}
&s_0(t) =& G^Y \begin{bmatrix} p^Y(t)-p^Y_{d}(t)-\tilde{y}^Y(t) \\ q^Y(t)-q^Y_d(t) \end{bmatrix}+\nonumber \\
&&G^\Delta  \begin{bmatrix} p^\Delta(t)-p^\Delta_{d}(t)-\tilde{y}^\Delta(t) \\ q^\Delta(t)-q^\Delta_d(t) \end{bmatrix}  +c, \forall t \in \mathscr{T}, \qquad
\label{eq6}
\end{eqnarray}
\noindent where $\tilde{y}^Y(t) \in \mathbb{R}^{3N_\text{bus}}$ is vector that has zero entries for buses and phases that do not have an EV, and is $y_j(t)$ for bus $n_j$ and phase $\phi_j$ with a wye connection as determined by the triplet $\mathscr{H}_j = \{n_j,\phi_j, \xi_j\}$ as defined in Section~\ref{sec:EVmod}. Similarly we may define $\tilde{y}^\Delta(t) \in \mathbb{R}^{3N_\text{bus}}$ for delta connection. Equation \eqref{eq6} represents two equations, one for the real and one for reactive component. The voltage magnitude constraints are denoted by
\begin{equation}
\varv_n^{\phi,\text{min}} \leq \varv_n^\phi(t) \leq \varv_n^{\phi,\text{max}}, \forall n \in \mathscr{N}_\text{bus}, \phi \in \Phi, \forall t \in \mathscr{T} .
\label{eq12}
\end{equation}
For both wye and delta connections the real and reactive power injections by DG are formulated as:
\begin{eqnarray}
p_n^{\phi,\text{min}} \leq p_n^\phi(t) \leq p_n^{\phi,\text{max}}, \forall n \in \mathscr{N}_\text{gen}, \phi \in \Phi, \forall t \in \mathscr{T},  \\
q_n^{\phi,\text{min}} \leq q_n^\phi(t)  \leq q_n^{\phi,\text{max}}, \forall n \in \mathscr{N}_\text{gen}, \phi \in \Phi, \forall t \in \mathscr{T},
\label{eq11}
\end{eqnarray}
\noindent where $\mathscr{N}_\text{gen} \subseteq \mathscr{N}_\text{bus}$ is the set of nodes that contain DG. For $n \in N_\text{bus}/\mathscr{N}_\text{gen} $ we have $p_n^\phi = q_n^\phi = 0$ for all $\phi \in \Phi$. For $n \in \mathscr{N}_\text{gen}$, if DG is connected to only one phase $p_n^\phi = q_n^\phi = 0$ for the remaining phases. The EV charging related constraints given in \eqref{eq2}, \eqref{eq1} that describe the intertemporal state of charge dynamics, non-negativity and charging rate constraints are also included.

The objectives of the DADM refer to the minimisation of the cost of real power procured at the substation:
\begin{equation}
\sum_{t \in \mathscr{T}} \sum_{\phi \in \Phi}\lambda_0(t) p^\phi_0(t) \Delta t,
\label{eq7}
\end{equation}
\noindent where $ \lambda_0(t)$ is the locational marginal price (LMP) at the substation at time $t$ and $p^\phi_0(t)$ is the injection at phase $\phi$ at time $t$ at the substation and $\Delta t$ is the time interval that the DADM is cleared, e.g., 5 minutes. A byproduct of \eqref{eq7} is that each EV $j \in \mathscr{E}$ procures the desired energy $e_j$ at minimum cost, as stated in Section~\ref{sec:EVmod}. The objective also includes a term that ensures that voltage levels throughout the network are operating close to the reference voltage: 
\begin{equation}
\sum_{t \in \mathscr{T}} \sum_{n \in \mathscr{N}_\text{bus}}\sum_{\phi \in \Phi} (\varv_n^\phi(t) - \varv_\text{ref})^2,
\label{eq8}
\end{equation}
\noindent where $\varv_\text{ref}$ is the reference voltage. The cost of distributed generation is also taken into account with 
\begin{equation}
\sum_{t \in \mathscr{T}}\sum_{n \in \mathscr{N}_\text{gen}} \sum_{\phi \in \Phi} c_n^\phi(t) p_n^\phi(t) \Delta t,
\label{eq9}
\end{equation}
\noindent where $c_n^\phi(t)$ is the cost of DG generation connected to node $n$ and phase $\phi$ at time $t$. The degradation cost of EV batteries given in \eqref{eq5} is also included in the formulation. The decision variables for each $n \in \mathscr{N}_\text{bus}$ and $\phi \in \Phi$ is the real power injection $p_n^\phi(t)$; the reactive power injection $q_n^\phi(t)$; the voltage magnitude $\varv_n^\phi(t)$; and for each EV $j \in \mathscr{E}$ is the charging schedule $y_j(t)$, for all $t \in \mathscr{T}$. The DADM is formulated as follows:

{\begin{align}
\min_{ \begin{subarray}{l} \{p_n^\phi(t), q_n^\phi(t),   \varv_n^\phi(t),y_j(t)\}\\ t \in \mathscr{T}, n \in \mathscr{N}_\text{bus}, \phi \in  \Phi , j \in \mathscr{E} \end{subarray}}  &  \eqref{eq5}+ \eqref{eq7}+\eqref{eq8}+\eqref{eq9} \nonumber \\
    \mathrm{subject~to~}& \eqref{eq2},\eqref{eq1},\eqref{eq10}-\eqref{eq11}
  \label{eq:dadm}
  \end{align}}

In the formulation above we only consider one-directional charging under the VGI framework. This can be easily extended to bi-directional charging. In this work we focus on one-directional charging as an intermediate step between the uncontrolled charging and V2G which requires a more intense communication network.

\section{Proposed Distributed DADM}

\label{sec:dadm_dis}

In this section, we formulate the distributed solution to the DADM that addresses privacy and computational issues of the market participants. The solution of \eqref{eq:dadm} by the DSO requires knowledge of EV information in terms of desired energy, departure and arrival times, and SOC. However, EV owners are reluctant in providing such information. Moreover, if the number of EVs is very large it can be very computationally intensive for the DSO to solve the DADM. In this regard, there is a need to propose a distributed solution of the DADM clearing mechanism. We divide the DADM participants into $E+1$ agents, i.e., the EV owners ($E$) and the DSO. The proposed framework could be extended to any number of agents, e.g., PV owners could also be separate agents or even each network bus; however, since the focus of this paper is on EV charging we limit the number of agents to $E+1$. We assume that the communication network that these agents use to exchange information is time-varying as is in reality. The DADM clearing mechanism given in \eqref{eq:dadm} may be seen as an optimisation problem where each agent optimises a local objective subject to local constraints, but needs to agree with the other agents in the network on the value of some decision variables that refer to the usage of shared resources, i.e., the power at the substation and the network usage, which are represented by coupling constraints. More specifically, each agent $i$ has its own vector $x_i \in \mathbb{R}^{n_i}$ of $n_i$ decision variables, e.g., the voltage magnitude, the charging schedule; its local linear constraint set $A_i x_i = b_i$ and $D_i x_i \leq 0$, these include constraints such as \eqref{eq2},\eqref{eq1}, \eqref{eq12}-\eqref{eq11}; and its objective $f_i(x_i): \mathbb{R}^{n_i} \rightarrow \mathbb{R}$, e.g., \eqref{eq5}, \eqref{eq7}-\eqref{eq9}. The coupling constraints refer to \eqref{eq10} and \eqref{eq6}; \eqref{eq10} has $3 N_\text{bus} T$ constraints and \eqref{eq6} has $6 T$ (since \eqref{eq6} refers to two equality constraints per time step) thus in total the coupling constraints are $3T(N_\text{bus}+2)$. We denote the  coupling constraints as $\sum_{i = 1}^{E+1} Z_i x_i = \zeta$, where $Z_i \in \mathbb{R}^{3T(N_\text{bus}+2) \times n_i}$ and $\zeta \in \mathbb{R}^{3T(N_\text{bus}+2)}$. Each agent contributes to the coupling constraints with $Z_i$. Now we may rewrite \eqref{eq:dadm} in compact form as
{\begin{align}
\min_{  \{x_i\}_{i = 1}^{E+1}}  &  \sum_{i =1}^{E+1}f_i(x_i) \nonumber \\
    \mathrm{subject~to~}& A_i x_i = b_i, i = 2,\dots,E+1,\nonumber \\
    &D_i x_i \leq 0 ,i = 1,\dots,E+1,\nonumber \\
    & \sum_{i = 1}^{E+1} Z_i x_i = \zeta.
  \label{eq:dadm1}
  \end{align}}

The construction and definitions of all variables and parameters, e.g., $x_i$ or $A_i$ may be found in the Appendix to facilitate the readability of the paper.

\subsection{Proposed distributed algorithm}

A distributed strategy that addresses both privacy and computational issues of the DADM given in \eqref{eq:dadm1} is:

\begin{algorithm}[h!]
\caption{Distributed DADM}
\begin{spacing}{1}
\begin{algorithmic}[1]
\STATE \textbf{Initialization} \\
\STATE ~~~~$k=0$. \\
\STATE ~~~~Consider $\xhi(0)$ such that $A_i \xhi(0) = b_i, \, D_i \xhi(0) \leq 0$, \\~~~~for all $i=1,\dots,E+1$. \\
\STATE ~~~~Consider $\li(0)\in\bR^r$, for all $i=1,\dots,E+1$. \\
\STATE ~~~~\textbf{For $i=1,\ldots,E+1$ repeat until convergence} \\
\STATE ~~~~$\aik = \sum_{j=1}^{E+1} a_j^i(k) \ljk$. \\
\STATE ~~~~$X_i = \{x_i: A_ix_i = b_i, \, D_i x_i \leq 0\}$ \\ ~~~~$\xikp \in \arg\min_{x_i \in X_i} f_i(x_i) + \aik\T Z_i x_i$. \\
% \STATE ~~~~$\likp = [\aik + c(k) g_i(\xikp)]^+$.
\STATE ~~~~$\likp = \aik+c(k)(Z_i \xikp-\frac{\zeta}{E+1})$
\STATE ~~~~$\xhikp = \xhik + \frac{c(k)}{\sum_{r=0}^k c(r)} (\xikp-\xhik)$.
\STATE~~~~$\tilde{x}_i(k+1) = \begin{cases}
    \xhikp &, k< k_{i,s}\\
   \frac{\sum_{r=k_{i,s}}^k c(r) \xikp}{\sum_{r=k_{i,s}}^k c(r)}  &, k \geq k_{i,s}
\end{cases}$.
\STATE ~~~~$k \gets k+1$.
\end{algorithmic}
\end{spacing}
\label{alg:Alg1}
\end{algorithm}

In the Algorithm above $r$ is the row dimension of the $Z_i$ matrices, i.e., the number of coupling constraints which are $r = 3T(N_\text{bus}+2)$, $c(k)$ is the subgradient step-size usually set to $c(k) = \frac{\beta}{k+1}$ for some $\beta>0$, $k_{s,i} \in \mathbb{N}_+$ is the iteration index related to a specific event, namely, the convergence of the Lagrange multipliers, as detected by agent $i$.

The steps of the algorithm may be explained as follows. Each agent $i$, $i = 1,\dots,E+1$ initialises the estimate of its local decision vector with $\xhi(0)$ that needs to satisfy its local constraints, i.e., $\xhi(0)$ such that $A_i\xhi(0) = b_i, \, D_i \xhi(0) \leq 0$ (step 3 of algorithm), and the estimate of the common dual variables vector of the equality constraints with a $\li(0)\in\bR^r$, e.g., $\li(0) = 0_{r \times 1}, i = 1,\dots, E+1$. At every iteration $k$ each agent $i$ computes a weighted average $\aik$ of dual variables vector based on the estimates $\ljk$, $j = 1,\dots,E+1$, of the other agents and its own estimate. The weight $a_j^i(k)$ that agent $i$ attributes to the estimate of agent $j$ at iteration $k$ is set equal to zero if agent i does not communicate with agent $j$ at iteration $k$. The conditions that the communication network weights must satisfy are the following: $a_j^i(k) \in [0,1)$, for all $k \geq 0$, $\sum_{j=1}^{E+1} a_j^i(k) = 1$, $\forall i = 1,\dots, E+1$, $\sum_{i=1}^{E+1} a_j^i(k) = 1$, $\forall j = 1,\dots, E+1$. Agent $i$ updates its local variable $\tilde{x}_i$ until convergence.

Some important characteristics of the algorithm are that no local information related to the primal problem is exchanged between the agents. In particular, only the estimates of the dual vector are communicated; thus addressing privacy concerns of the DSO and the EV owners. Furthermore, the algorithm reduces computational complexity by distributing the burden between the agents. The communication network of the DSO and the EV owners may be time-varying and has to satisfy the constraints mentioned above for $a_j^i(k)$. Step 9 of the algorithm is a running average of the primal iterates which are constructed as they are shown to exhibit superior convergence properties with respect to $x_i(k)$ while step 10 performs a reset of this average at a certain iteration index as this has been shown to speed up practical convergence~\cite{agapoua}. It has been shown that the dual iterates $\li(k)$ generated by the algorithm converge to an optimal dual vector which $\xhi(k)$ achieve asymptotically the optimal objective value. More details about the algorithm may be found in~\cite{agapoua}.

\section{Uncertainty modelling}

\label{sec:unc_mod}

In this section we propose a methodology to provide the EV aggregators with a given confidence on the amount of capacity that they have available to participate in the DADM. The proposed framework may be used by aggregators to provide services to the grid. Given that the EV owners do not share any private information they would be willing to let an aggregator be responsible to charge their vehicle subject to their desires, e.g., final state of charge, departure times, etc. for receiving monetary benefits. The aggregator would communicate with the DSO and other aggregators to determine the clearing of the DADM. One challenge in this setting is that the aggregator would not be certain of the number of EVs available at a certain time instant, their state of charge and available energy, their arrival and departure times. 

In order for the aggregators to have a given confidence on the amount of capacity that they have available to participate in the market we use a simulation approach that contains independent Monte Carlo simulations and requires  the construction of multiple independent and identically distributed (i.i.d.) sample paths for each output random variable to evaluate the performance metrics~\cite[p.~10]{monte}. More specifically, we carry out simulation runs to determine the probability distribution of the amount of available capacity by the aggregator for every time interval, i.e., $\gamma(t) = \sum_{j \in \mathscr{E}} y_j(t)$. The performance metrics we select is the expected value. Let $M$ be the number of Monte Carlo simulations, the estimated average available capacity by the aggregator for every time interval, i.e.,  
\begin{equation}
\overline{\gamma}(t) = \frac{1}{M} \sum_{m = 1}^M \gamma^{(m)}(t),
\end{equation}
\noindent where $\gamma^{(m)}(t)$ is the realization of the random variable $\gamma(t)$ in simulation run $m$. The number of simulation runs $M$ depends on the statistical reliability requirements specified for the estimation of the desired expected values. We define the statistical reliability of the hourly sample mean estimator $\overline{\gamma}(t)$ to be the length of the $100(1-\nu)\%$ confidence interval with $0<\nu<1$ for the true mean of $\gamma(t)$. According to the Central Limit Theorem, the sample mean estimator $\overline{\gamma}(t)$ is approximately normally distributed for large $M$~\cite{sim}. Thus we can establish that the true mean of $\gamma(t)$ lies in the interval 
\begin{equation}\left[ \overline{\gamma}(t)-z_{1-\nu/2} \frac{\sigma_{\gamma(t)}}{\sqrt{M}},\overline{\gamma}(t)+z_{1-\nu/2} \frac{\sigma_{\gamma(t)}}{\sqrt{M}} \right],
\end{equation}
\noindent with a $100(1-\nu)\%$ probability, where $\sigma_{\gamma}(t)$ is the standard deviation of $\gamma(t)$ and $z_{1-\nu/2} = \Phi^{-1}(1-\nu)$, with $\Phi^{-1}$ the inverse of the cumulative distribution of the standard normal distribution. The length of the confidence interval is a function of $\sqrt{M}^{-1}$ with decays slowly for large $M$. Thus beyond a certain value of $M$, the improvement in statistical reliability is generally too small to warrant the extra computing time needed to perform additional simulation runs. 

The confidence interval may be further tuned if more refined historical data are used to construct an empirical pdf, so that the aggregator may make a more informed decision. For instance, if the aggregator holds data for the location of the feeder and it is part of a rural or urban area, then the predicted daily energy consumption of an EV may be tuned accordingly (e.g.,~\cite{8450584}).

\section{Numerical Results}

\label{sec:num}

In this section, we present several numerical examples to demonstrate the capabilities of the proposed VGI framework. We use small systems, the unbalanced 13-bus and 33-bus distribution feeders to provide insights into the results presented. We demonstrate the scalability of the proposed distributed algorithm in Section~\ref{sec:dadm_dis} with the 141-bus distribution feeder~\cite{5491276} with 11 agents communicating with each other. Additionally, we demonstrate how the amount of EVs affects the confidence level that the EV aggregator has when participating in the DADM.

\begin{figure}[t!]
     \centering
  \includegraphics[width=.44\textwidth]{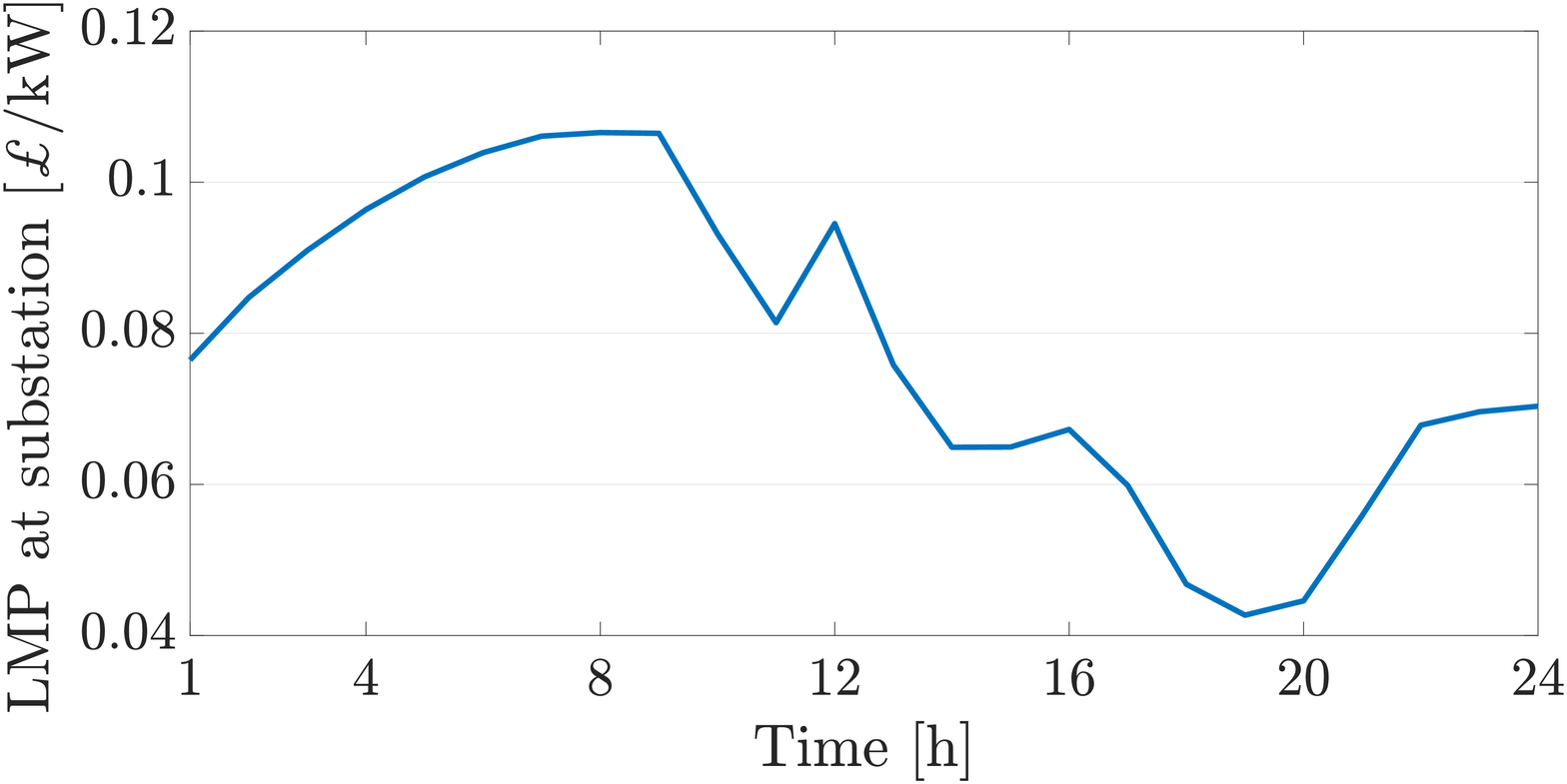}   
  \caption{LMP at the substation over a 24 hour period.}
     \label{fig2}
     \vspace{-\baselineskip}
 \end{figure}

\begin{figure}[b!]
\vspace{-\baselineskip}
     \centering
  \includegraphics[width=.44\textwidth]{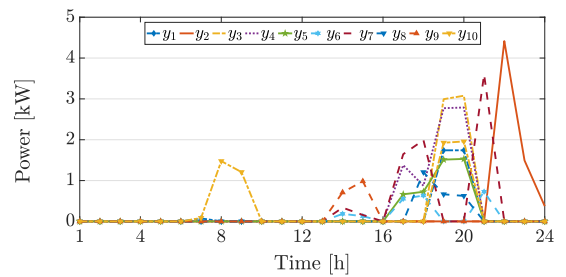}   
  \caption{Charging schedule of 10 EVs over a 24 hour period.}
     \label{fig3}
 \end{figure}

\begin{figure}[t!]
     \centering
  \includegraphics[width=.44\textwidth]{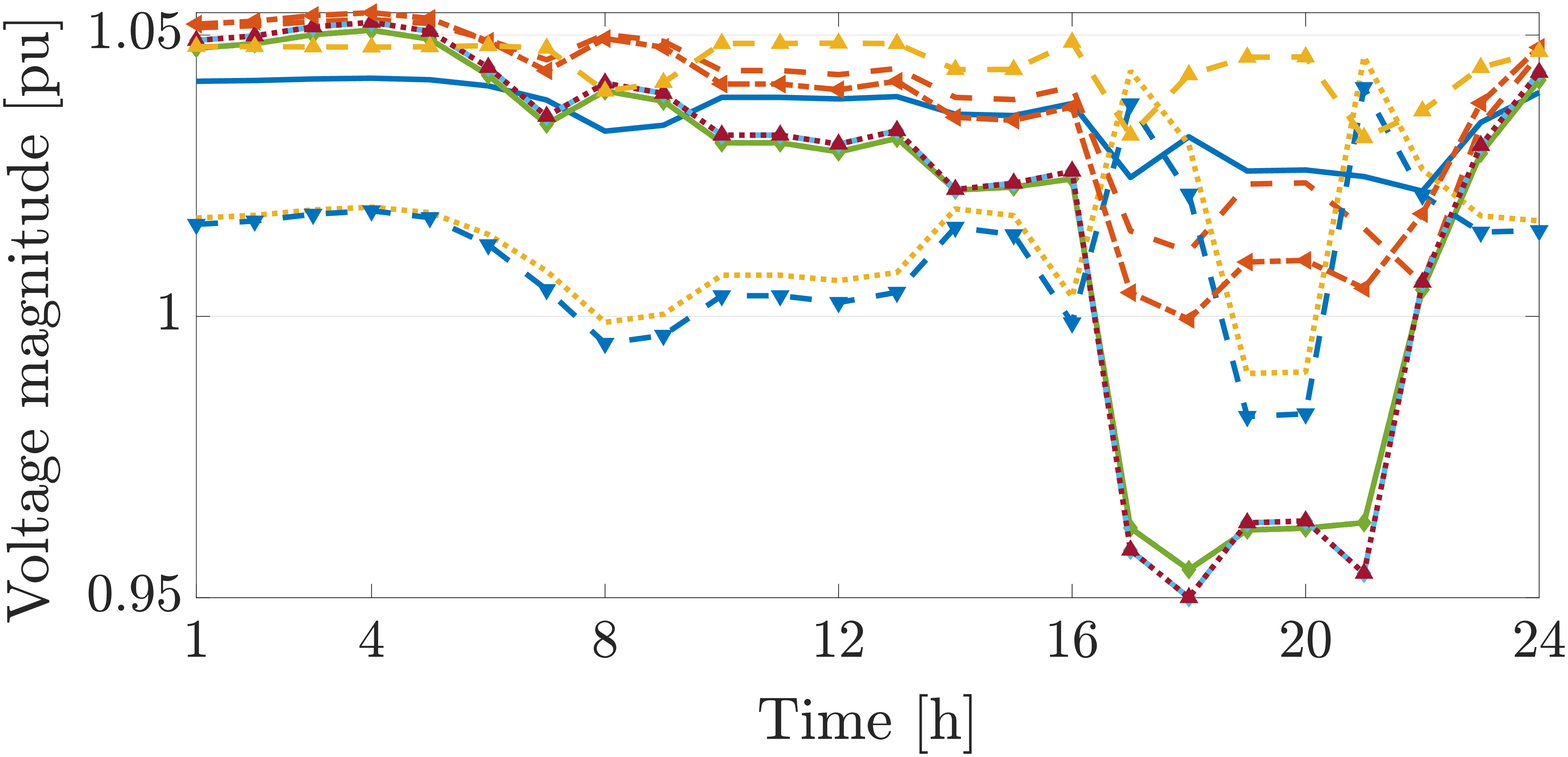}   
  \caption{Voltage magnitude at the EVs connections of the 13-bus feeder over a 24 hour period.}
     \label{fig4}
     \vspace{-\baselineskip}
 \end{figure}

%
%  \begin{figure}[t!]
%     \centering
%  \includegraphics[width=.44\textwidth]{fig1.eps}   
%  \caption{The probability an EV is at home with time on a weekday.}
%     \label{fig1}
% \end{figure}

\subsection{13-bus distribution feeder}

A lot of work has conducted into obtaining realistic data for EV behaviour (see, e.g.,~\cite{8450584}, ~\cite{con1},~\cite{ev_national_grid}). In this work, we use data from~\cite{8450584} to model realistically the charging behaviour of an EV. In the numerical results we assume that charging is taking place at home. 

%
%The distribution of the predicted daily energy consumption in kWh may be found in Table~\ref{tab:ev} and the probability an EV is at home with time is depicted in Fig~\ref{fig1}. 
%
%\begin{table}[t!]
%\begin{center}
%\resizebox{0.47\textwidth}{!}{
%\begin{tabular}{|c|c|c|c|c|c|c|c|} 
% \hline
%  0 & 0-5 & 5-10 & 10-20 & 20-30 & 30-50 & 50-100 & $>$100\\ 
% \hline
%27.0\% & 38.6\% &17.6\% & 9.0\% & 3.8\% & 2.4\% & 1.4\% & 0.2\% \\
% \hline
%\end{tabular}}
%\end{center}
%\caption{The distribution of predicted daily energy consumption in kWh.}
%\label{tab:ev}
%\end{table}

In the first case study we consider the 13-bus distribution feeder~\cite{119237} with no renewable resources. We consider a collection of 10 EVs denoted by the set $\mathscr{E} =\{1,2,...,10\}$ that are wye connected in various phases of the 13-bus feeder and a study period of $\mathscr{T} = \{1,...,24\}$ with intervals of size $\Delta t = 1$ h. For instance, EV 1 is at node 632 in phace c with times of arrival and departure specified by $t_{1,1}^\text{dep} = 11$ and $t_{1,1}^\text{arr} = 12$ and required energy $e_1 = 3.56$ kWh. The maximum charging value is set to $y_j^{\text{}max} = 10$ kW for all $j = 1,\dots,10$. The LMP at the substation $\lambda_0(t)$ for $t = 1,\dots,24$ is depicted in Fig~\ref{fig2}. The minimum (maximum) allowed voltage level is 0.95 pu (1.06 pu). The outcome of the charging schedule of the EVs is depicted in Fig.~\ref{fig3}. As it may be seen the EVs that are available select to  charge at the hours when the LMP at the substation is lower, i.e., 16:00-22:00. At the same time interval we may notice in Fig.~\ref{fig4} that the voltage levels are near the minimum value for the nodes and phases where the EVs are connected. 

\subsection{33-bus distribution feeder}

\begin{figure}[b!]
\vspace{-\baselineskip}
     \centering
  \includegraphics[width=.44\textwidth]{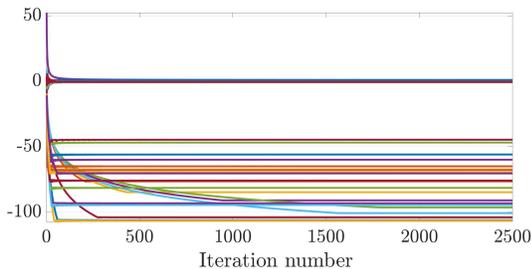}   
  \caption{Evolution of the agents’ estimates $\{\kappa_i(k)\}_{i = 1}^6$ where all agents communicate with each other for the 33-bus system.}
     \label{fig5}
 \end{figure}

\begin{figure}[t!]
     \centering
  \includegraphics[width=.44\textwidth]{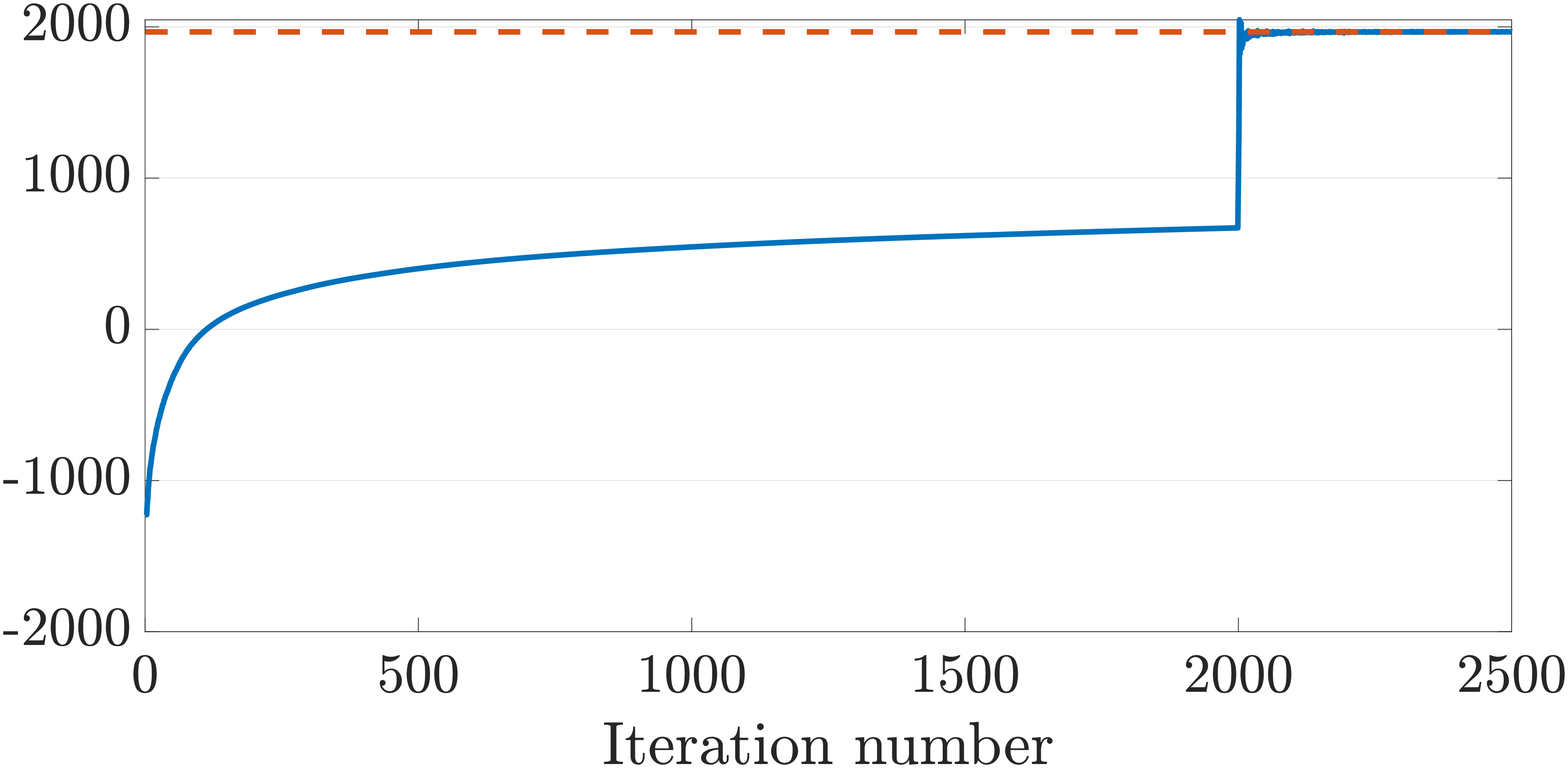}   
  \caption{Evolution of objective function until it reaches the optimal value (red line) where all agents communicate with each other for the 33-bus system.}
     \label{fig6}
     \vspace{-\baselineskip}
 \end{figure}

\begin{figure}[b!]
\vspace{-1.6\baselineskip}
     \centering
  \includegraphics[width=.15\textwidth]{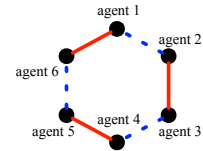}   
  \caption{Time-varying communication network where half the agents communicate with each other for the 33-bus system.}
     \label{figcom}
 \end{figure}
 
\begin{figure}[b!]
\vspace{-\baselineskip}
     \centering
  \includegraphics[width=.44\textwidth]{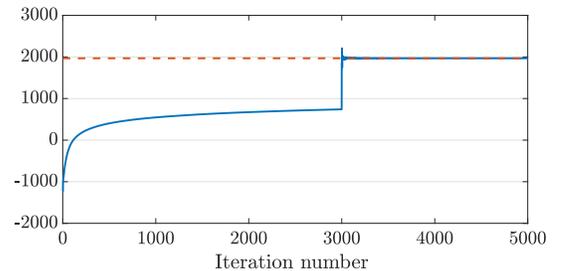}   
  \caption{Evolution of objective function until it reaches the optimal value (red line) where half the agents communicate with each other for the 33-bus system.}
     \label{fig7}
 \end{figure}

The solution of \eqref{eq:dadm} by the DSO requires knowledge of EV information in terms of desired energy, departure and arrival times, and SOC. However, EV owners are reluctant in providing such information. Moreover, if the number of EVs is very large it can be very computationally intensive for the DSO to solve the DADM. In this regard, we formulated a distributed solution of the DADM clearing mechanism given in Section~\ref{sec:dadm_dis}. We validate the proposed methodology in a 33-bus with study period of $\mathscr{T} = \{1,...,24\}$ with intervals of size $\Delta t = 1$ h. We divide the DADM participants into $6$ agents, i.e., the EV owners ($5$) and the DSO. The optimisation problem of the DSO has 2016 decision variables and local constraints set defined by 4032 inequalities. The optimisation problem of each EV has 96 decision variables and local constraints set defined by 1 equality and 192 inequalities. There are 840 coupling equality constraints, and therefore we have 840 Lagrange multipliers associated with them. It is assumed that all agents communicate with each other and the $c(k)=\frac{100}{k+1}$, $k_{i,s} = 2000$ for $i = 1,\dots, 6$. We ran the proposed algorithm for 2500 iterations with $\kappa_i(0) = 0$, $i = 1,\dots,6$ and the evolution of the Lagrange multipliers is depicted in Fig.~\ref{fig5}. As we may see they converge to the optimal value from around 2000 iterations. In Fig.~\ref{fig6} the evolution of the objective value is depicted. We may see a jump at iteration number $k_{i,s} = 2000$ for $i = 1,\dots, 6$ since the Lagrangian multipliers have converged and we only use estimates for $\hat{x}_i(k)$ based on values after iteration $k_{i,s}$ (see step 10 of the algorithm). In order to test how the communication network affects the rate of convergence we modify the communication network so that only half the agents talk with the other half at any time-step. The communication network is depicted in Fig.~\ref{figcom} and corresponds to a connected graph, whose edges are divided into two groups: the blue and the red ones, which are activated alternatively. Here, we have $k_{i,s} = 3000$ for $i = 1,\dots, 6$. According to step 10 of the proposed distributed algorithm the ``jump'' at iteration $k_{i,s}$ speeds up practical convergence by ``resetting'' the running average estimate. We may notice in Fig.~\ref{fig7} that the objective function now converges to the optimal value at a greater number of iterations compared to the previous case where all agents communicated to each other as seen in Fig.~\ref{fig6}.

\subsection{141-bus distribution feeder}

\begin{figure}[t!]
     \centering
  \includegraphics[width=.44\textwidth]{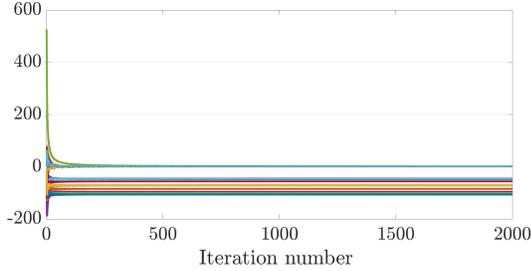}   
  \caption{Evolution of the agents’ estimates $\{\kappa_i(k)\}_{i = 1}^{11}$ where all agents communicate with each other for the 141-bus system.}
     \label{fig8}
     \vspace{-\baselineskip}
 \end{figure}

\begin{figure}[b!]
\vspace{-\baselineskip}
     \centering
  \includegraphics[width=.44\textwidth]{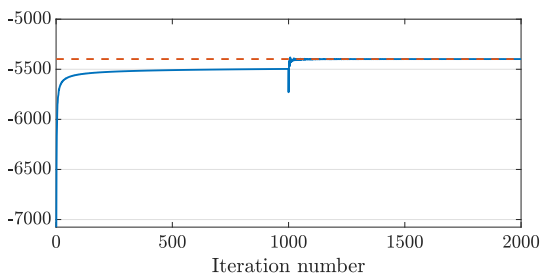}   
  \caption{Evolution of objective function until it reaches the optimal value (red line) for the 141-bus system.}
     \label{fig9}
 \end{figure}

To demonstrate the scalability of the proposed methodology we implement the distributed algorithm in the 141-bus distribution feeder. In this case the DADM has 11 agents, i.e., the EV owners (10) and the DSO. The optimisation problem of the DSO has 9408 decision variables and local constraints set defined by 18816 inequalities. The optimisation problem of each EV has 96 decision variables and local constraints set defined by 1 equality and 192 inequalities. There are 3408 coupling equality constraints, and therefore we have 3408 Lagrange multipliers associated with them. It is assumed that all agents communicate with each other and the $c(k)=\frac{1000}{k+1}$, $k_{i,s} = 1000$ for $i = 1,\dots, 11$. We ran the proposed algorithm for 2000 iterations with $\kappa_i(0) = 0$, $i = 1,\dots,11$ and the evolution of the Lagrange multipliers is depicted in Fig.~\ref{fig8}. As we may see they converge to the optimal value from around 1000 iterations. In Fig.~\ref{fig9} the evolution of the objective value is depicted. We may see a jump at iteration number $k_{i,s} = 1000$ for $i = 1,\dots, 11$ since the Lagrangian multipliers have converged and we only use estimates for $\hat{x}_i(k)$ based on values after iteration $k_{i,s}$.

\begin{figure}[t!]
     \centering
  \includegraphics[width=.44\textwidth]{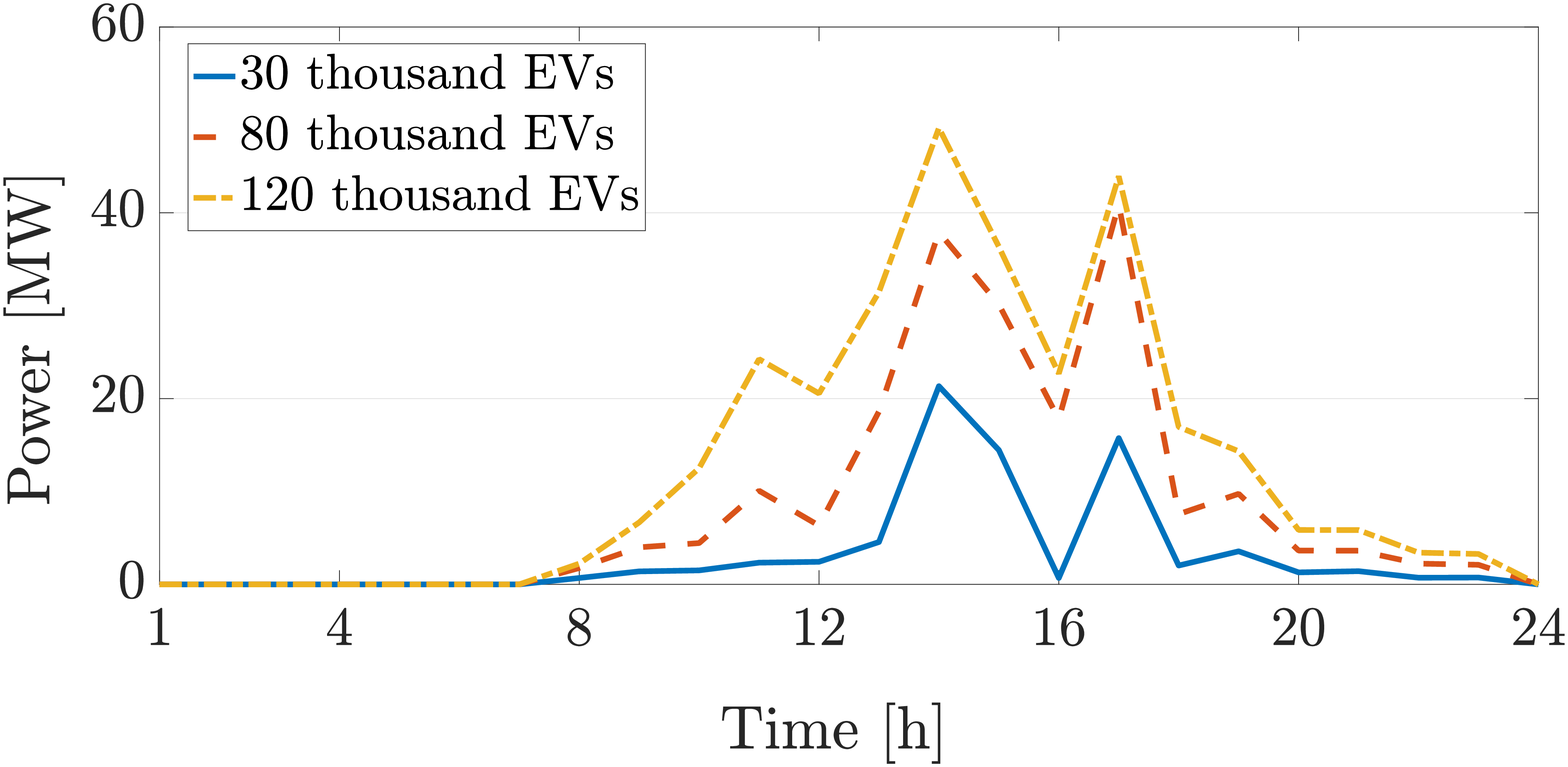}   
  \caption{Hourly mean values of available capacity for the EV aggregator with 95\% confidence for various levels of EV integration for the 141-bus system.}
     \label{fig10}
%     \vspace{-\baselineskip}
 \end{figure}

\begin{figure}[b!]
\vspace{-\baselineskip}
     \centering
  \includegraphics[width=.44\textwidth]{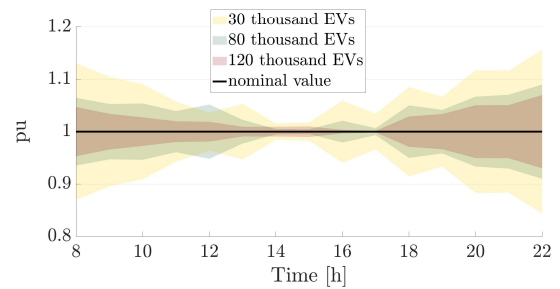}   
  \caption{Normalised mean value intervals with 95\% confidence for the 141-bus system.}
     \label{fig11}
 \end{figure}

In order to perform an analysis on the number of vehicles that are necessary for an aggregator to participate in the DADM with a confidence interval we perform Monte Carlo simulations in the 141-bus feeder. We modify the level of EV deployment from: i) 30, ii) 80, and iii) 120 thousand EVs and run 500 Monte Carlo simulations for each integration level. In Fig.~\ref{fig10}, we depict the mean value of the available capacity for every hour of the day; we may notice that as the number of EVs increases capacity is also available at more hours of the day. In Fig.~\ref{fig11}, we depict the intervals with 95\% confidence for these mean values, the values are normalised with the mean value for every level of integration so that the graph is more readable. We may notice that as the number of EVs increases the level of confidence that the EV aggregator has also increases. In Table~\ref{tab:evunc} we show the available capacity for hour $t=14$ as a function of various confidence intervals and levels of EV integration.

\begin{table}[t!]
\begin{center}
\resizebox{0.47\textwidth}{!}{\begin{tabular}{|c|c|c|c|c|} 
 \hline
 \# thousands &  Mean Value of  & \multicolumn{3}{c|}{Confidence interval [MW]} \\
    \cline{3-5} 
    of EVs & Available capacity [MW] & 99\%  & 95\%  & 90\%  \\
 \hline
 30 &21.1&[20.9, 21.8]&[21.0, 21.7] & [21.1, 21.6] \\
  \hline 
  80 &37.9 &[37.7, 38.2]&[37.7, 38.1] &[37.8, 38.1] \\
  \hline 
  120&49.1&[48.5, 49.8]&[48.7, 49.6] &[48.8, 49.5] \\
  \hline 
\end{tabular}}
\end{center}
\caption{EV aggregator confidence level as a function of EVs for hour $t=14$.}
\label{tab:evunc}
\vspace{-\baselineskip}
\end{table}

\section{Conclusions}

\label{sec:conc}

In this paper, we developed a distributed VGI framework that enables the smooth integration of EVs. Through the numerical examples, we demonstrated that the proposed framework is scalable and performs well in a variety of circumstances. We also demonstrated that this framework is useful for EV aggregators. More specifically, we provided a detailed model of EVs, i.e., representing their times of arrival and departure, SOC, required energy and objectives. Next, we formulated the centralised DADM that incorporates the charging needs of EVs and has a detailed representation of the underlying three phase power network. We proposed a distributed solution to the DADM that converges to the optimal solution under a time-varying communication network with no exchange of sensitive information. We provided the EV aggregator with a methodology to quantify the level of confidence for the available capacity that can participate in the DADM based on the number of EVs available.  

There are natural extensions of the work presented here. For instance, we will investigate the incentives for an EV owner to participate in this market setup. In our future studies, we plan on incorporating uncertainty in the formulation of the DADM due to load variations and the intermittent nature of renewable resources and propose a distributed algorithm that converges under this uncertain environment. We will report on these developments in future papers.

\appendix

The three-phase network studied has $E$ EVs connected to it and a total of $3 N_\text{bus}$ connection points. Thus $3 N_\text{bus}-E$ are nodes/phases that do not contain an EV and are part of the DSO agent. The DSO agent, which without loss of generality, is indexed by $1$, has a decision variable $x_1 \in \mathbb{R}^{n_1}$ with $n_1= 3 (3 N_\text{bus}-E)T$. More specifically:
\begin{equation*}
\resizebox{0.95\hsize}{!}{%
        $x_1 = [p_n^\phi(t), q_n^\phi(t),\varv_n^\phi(t): \{n,\phi,\xi\} \notin \mathscr{H}, n \in \mathscr{N}_\text{bus}, \phi \in \Phi, t \in \mathscr{T}]^\top$}.
\end{equation*}
\noindent Its objective is defined as $f_1 = \sum_{\begin{subarray}{l} \{n,\phi,\xi\} \notin \mathscr{H} \\ n \in \mathscr{N}_\text{bus}, \phi \in  \Phi \\ t \in \mathscr{T} \end{subarray}}  f_n^\phi(t)$, where 
\begin{equation}
f_n^\phi(t)= \left\{ \begin{matrix} (\varv_n^\phi(t) - \varv_\text{ref})^2+c_n^\phi(t) p_n^\phi(t) , & n\in \mathscr{N}_\text{gen} \\
(\varv_n^\phi(t) - \varv_\text{ref})^2, & n \in  \mathscr{N}_\text{bus} / \mathscr{N}_\text{gen} \cup \{0\} \\
\lambda_0(t) p^\phi_0(t)+(\varv_0^\phi(t) - \varv_\text{ref})^2, & n = 0
\end{matrix} \right.
\end{equation}

The limiting constraints for agent 1 given in \eqref{eq12}-\eqref{eq11} may be represented as the matrix 
\begin{equation}
D_1 = [C_{n}^\phi (t): \{n,\phi,\xi\} \notin \mathscr{H}, n \in \mathscr{N}_\text{bus}, \phi \in \Phi, t \in \mathscr{T}]^\top,
\end{equation}
\noindent where $C_{n}^\phi \in \mathbb{R}^{6 n_1 \times n_1}$, i.e., one row for the minimum and another for the maximum limit associated with each of the three variables.  

For agent $i: i=2,..\dots, E+1$, i.e., an EV owner $j$ connected to node $n_j$ and phase $\phi_j$ determined by the duplet $\{n_j,\phi_j,\xi_j\} \in \mathscr{H}_j$, we define the vector $x_i \in \mathbb{R}^{n_i}$ with $n_i = 4T$ by 
\begin{equation}
x_i = [p_{n_j}^{\phi_j}(t), q_{n_j}^{\phi_j}(t),\varv_{n_j}^{\phi_j}(t),y_j(t): t \in \mathscr{T}]^\top,
\end{equation}
and the objective function
\begin{equation}
f_i =  \sum_{t \in \mathscr{T}}\left((\varv_{n_j}^{\phi_j}(t) - \varv_\text{ref})^2+y^2_j(t)\right) .
\end{equation}

\noindent We rewrite \eqref{eq2} as  $A_i  x_i  = b_i$, where $A_{i} \in \mathbb{R}^{1 \times n_i}$ and $b_{i} \in \mathbb{R}$ and the limiting constraints given in \eqref{eq1} and \eqref{eq12}-\eqref{eq11} as  $D_i = [C_{n_j}^{\phi_j}(t): t \in \mathscr{T}]^\top$, where $C_{n_j}^{\phi_j} \in \mathbb{R}^{8 \times n_i}$, i.e., one row for the minimum and another for the maximum limit associated with each of the four variables.

\bibliographystyle{IEEEtran}
\bibliography{journals-full,references}

\balance

\end{document}